\documentclass{article}

\usepackage{PRIMEarxiv}

\usepackage[utf8]{inputenc} 
\usepackage[T1]{fontenc}    
\usepackage{hyperref}       
\usepackage{url}            
\usepackage{booktabs}       
\usepackage{amsfonts}       
\usepackage{nicefrac}       
\usepackage{microtype}      
\usepackage{lipsum}
\usepackage{fancyhdr}       
\usepackage{graphicx}       
\graphicspath{{media/}}     

\usepackage{amsmath,amssymb,amsfonts}
\usepackage{algorithmic}
\usepackage{textcomp}
\usepackage{xcolor}
\usepackage{makecell}
\usepackage{multirow}

\usepackage{bm}
\usepackage{adjustbox}

\title{Blind Video Quality Assessment at the Edge}

\author{
  Zhanxuan Mei \\
  University of Southern California \\
  Los Angeles, USA\\
  \texttt{zhanxuan@usc.edu} \\
   \And
  Yun-Cheng Wang \\
  University of Southern California \\
  Los Angeles, USA\\
  \texttt{yunchenw@@usc.edu} \\
   \And
  C.-C. Jay Kuo \\
  University of Southern California \\
  Los Angeles, USA\\
  \texttt{cckuo@sipi.usc.edu} \\
}

\begin{document}
\maketitle

\begin{abstract}
Owing to the proliferation of user-generated videos on the Internet,
blind video quality assessment (BVQA) at the edge attracts growing
attention.  The usage of deep-learning-based methods is restricted to be
applied at the edge due to their large model sizes and high
computational complexity.  In light of this, a novel lightweight BVQA
method called GreenBVQA is proposed in this work. GreenBVQA features a
small model size, low computational complexity, and high performance.
Its processing pipeline includes: video data cropping, unsupervised
representation generation, supervised feature selection, and
mean-opinion-score (MOS) regression and ensembles. We conduct
experimental evaluations on three BVQA datasets and show that GreenBVQA
can offer state-of-the-art performance in PLCC and SROCC metrics while
demanding significantly smaller model sizes and lower computational
complexity. Thus, GreenBVQA is well-suited for edge devices. 
\end{abstract}


\section{Introduction}\label{G_sec:introduction}

Objective video quality assessment methods are often
classified into three categories: full-reference video quality
assessment (FR-VQA), reduced-reference video quality assessment
(RR-VQA), and no-reference video quality assessment (NR-VQA).  NR-VQA is
also known as blind video quality assessment (BVQA). FR-VQA methods
assess video quality by measuring the difference between distorted
videos and their reference videos. One well-known example is VMAF
\cite{li2016toward, lin2014fusion}.  RR-VQA
\cite{soundararajan2012video} methods evaluate video quality by
utilizing a part of information from reference videos, which offers
greater flexibility than FR-VQA. Finally, BVQA is the only choice if
there is no reference video available.  With the rise of social media
and the popularity of multi-party video conferencing, there has been an
explosion of user-generated video content. A significant portion of the
user-generated content lacks the availability of reference videos,
necessitating the need for BVQA methods to automatically and efficiently
evaluate perceptual video quality.  Furthermore, the adoption of edge
computing is on the rise, primarily attributable to its capacity to
reduce latency, conserve network bandwidth, enhance privacy, and enable
real-time data processing. In edge computing environments, BVQA plays a
pivotal role in ensuring that video quality remains high while
optimizing resources and responsiveness.  Thus, BVQA attracts growing
attention in recent years, as it addresses the pressing need of
evaluating video quality in diverse contexts, spanning user-generated
content and edge computing scenarios. 

One straightforward BVQA solution is to build it upon blind image
quality assessment (BIQA) methods. That is, the application of BIQA
methods to a set of key frames of distorted videos individually.  BIQA
methods can be classified into three categories: natural scene statistic
(NSS) based methods \cite{mittal2012making, moorthy2011blind,
mittal2012no}, codebook-based methods \cite{ye2012unsupervised,
xu2016blind} and deep-learning-based (DL-based) methods
\cite{bosse2017deep, zhang2018blind}.  However, directly applying BIQA
followed by frame-score aggregation does not produce satisfactory
outcomes because of the absence of temporal information. Thus, it is
essential to incorporate temporal or spatio-temporal information.  Other
BVQA methods with handcrafted features \cite{valenzise2011no,
sogaard2015no} were evaluated on synthetic-distortion datasets with
simulated distortions such as transmission and compression.  Recently,
they were evaluated on authentic-distortion datasets as well as reported
in \cite{saad2014blind, mittal2015completely}. Their performance on
authentic-distortion datasets is somehow limited. Authentic-distortion
VQA datasets arise from the user-generated content (UGC) in the real-world
environment. They contain complicated and mixed distortions with
highly diverse contents, devices, and capture conditions. 

Deep learning (DL) methods have been developed for BIQA and BVQA
\cite{zhang2018blind}. To further enhance performance and reduce
distributional shifts \cite{liu2021spatiotemporal}, pre-trained models
on large-scale image datasets, such as the ImageNet
\cite{deng2009imagenet}, are adopted in \cite{li2021unified,
tu2021rapique}.  However, it is expensive to adopt large pre-trained
models on mobile or edge devices.  Edge computing is a rapidly growing
field in recent years due to the popularity of smartphones and Internet
of Things (IoT).  It involves processing and analyzing data near their
source, typically at the ``edge'' of the network (rather than
transmitting them to a centralized location such as a cloud data
center). Given that a high volume of videos triggers heavy Internet
traffic, video processing at the edge reduces the video transmission
burden and saves the network bandwidth.  The VQA task can enhance many
video processing modules, such as video bitrate adaptation
\cite{bentaleb2018survey}, video quality assurance
\cite{safari2022edge}, and video pre-processing \cite{liu2021light}.
The demand for high-quality videos is increasing on edge devices, while
most user-generated contents lack reference videos, necessitating the
use of BVQA methods.  Furthermore, DL-based BVQA methods are expensive
to deploy on edge devices due to their high computational complexity and
large model sizes. 

Therefore, a lightweight BVQA method is demanded at the edge.  Based on
the green learning principle \cite{kuo2022green, kuo2016understanding},
a lightweight BIQA method, called GreenBIQA, was proposed in
\cite{mei2022greenbiqa} recently.  It is worth noting that GreenBIQA
exhibits limited performance when applied to VQA datasets, primarily due
to its lack of temporal information integration. Moreover, direct
deployment of GreenBIQA in video quality evaluation is computationally
expensive because of a huge difference in image and video data sizes.
To address this void, we propose a lightweight BVQA method and call it
GreenBVQA in this work. GreenBVQA features a smaller model size and
lower computational complexity while achieving highly competitive
performance against state-of-the-art DL methods. The processing pipeline
of GreenBVQA contains four modules: 1) video data cropping, 2)
unsupervised representation generation, 3) supervised feature selection,
and 4) mean-opinion-score (MOS) regression and ensembles.  The video
data cropping operation in Module 1 is a pre-processing step.  Then, we
extract spatial, spatio-color, temporal, and spatio-temporal
representations from cropped data in an unsupervised manner to obtain a
rich set of representations at low complexity in Module 2 and select a
subset of the most relevant features using the relevant feature test
(RFT) \cite{yang2022supervised} in Module 3. Finally, all selected
features are concatenated and fed to a trained MOS regressor to predict
multiple video quality scores and, then, an ensemble scheme is used to
aggregate multiple regression scores into one ultimate score.  
conduct experimental evaluations on three VQA datasets and show that
GreenBVQA can offer state-of-the-art performance in PLCC and SROCC
metrics while demanding significantly small model sizes, short inference
time, and low computational complexity. 

There are three main contributions of this work.
\begin{itemize}
\item A novel lightweight BVQA method, named GreenBVQA, is proposed. Four different types of
representation (i.e., spatial, spatio-color, temporal, and
spatio-temporal representations) are considered jointly.  Each type of
representation is passed to the supervised feature selection module
for dimension reduction. Then, all selected features are concatenated to
form the final feature set. 
\item Experiments are conducted on three commonly used VQA datasets to
demonstrate the advantages of the proposed GreenBVQA method.  Our method
outperforms all conventional BVQA methods in terms of MOS prediction
accuracy.  Its performance is highly competitive against DL-based
methods while featuring a significantly smaller model size, shorter
inference time, and lower computational complexity. 
\item A video-based edge computing system is presented to illustrate the
role of GreenBVQA in facilitating various video processing tasks at the
edge. The inherent characteristics of GreenBVQA, notably its lightweight
model and low computational complexity, serve as a compelling evidence
of its prospective utility in the realm of edge computing. 
\end{itemize} 

The rest of this paper is organized as follows. Related work is reviewed
in Sec. \ref{sec:related}.  The proposed GreenBVQA is presented in Sec.
\ref{sec:BVQA_method}. Experimental results are shown in Sec.
\ref{sec:experiments}.  An edge computing system employing GreenBVQA is
presented to demonstrate its potential real-world applications in Sec.
\ref{sec:edge_computing}.  Finally, concluding remarks are given and
future research directions are pointed out in Sec. \ref{sec:conclusion}.

\section{Review of Related Work}\label{sec:related}

Quite a few BVQA methods have been proposed in the last two decades.
Existing work can be classified into conventional and DL-based methods
two categories. They are first reviewed in Sec. \ref{subsec:CBVQA} and sec.
\ref{subsec:DLBVQA}, respectively. Then, we examine previous work on
video edge computing in Sec. \ref{subsec:VEC}

\subsection{Conventional BVQA Method}\label{subsec:CBVQA}

Conventional BVQA methods extract quality-related features from input
images using an ad hoc approach. Then, a regression model (e.g., Support
Vector Regression (SVR) \cite{awad2015support} or XGBoost
\cite{chen2016xgboost}) is trained to predict the quality score using
these handcrafted features.  One family of methods is built upon the
Natural Scene Statistics (NSS). NSS-based BIQA methods
\cite{moorthy2011blind, mittal2012no} can be extended to NSS-based BVQA
methods since videos are formed by multiple image frames. For example,
V-BLIINDS \cite{saad2012blind2} is an extension of a BIQA method by
incorporating a temporal model with motion coherency.  Spatio-temporal
NSS can be derived from a joint spatio-temporal domain.  For instance,
the method in \cite{li2016spatiotemporal} conducts the 3D discrete
cosine transform (DCT) and captures the spatio-temporal NSS of 3D-DCT
coefficients. The spatio-temporal statistics of mean-subtracted and
contrast-normalized (MSCN) coefficients of natural videos are
investigated in \cite{dendi2020no}, where an asymmetric generalized
Gaussian distribution (AGGD) is adopted to model the statistics of both
3D-MSCN coefficients and bandpass filter coefficients of natural videos.
Another family of BVQA methods, called codebook-based BVQA, is inspired
by CORNIA \cite{ye2012unsupervised}, \cite{xu2014no}. They first obtain
frame-level quality scores using unsupervised frame-based feature
extraction and supervised regression.  Then, they adopt temporal
pooling to derive the target video quality.  A two-level feature
extraction mechanism is employed by TLVQM \cite{korhonen2019two}, where
high- and low-level features are extracted from the whole sequence and a
subset of representative sequences, respectively. 

\subsection{DL-based BVQA Method}\label{subsec:DLBVQA}

DL-based BVQA methods have been investigated recently. They offer
state-of-the-art prediction performance.  Inspired by MEON
\cite{ma2017end}, V-MEON \cite{liu2018end} provides an end-to-end
learning framework by combining feature extraction and regression into a
single stage. It adopts a two-step training strategy: one for codec
classification and the other for quality score prediction.  COME
\cite{wang2018come} adopts CNNs to extract spatial features and uses
motion statistics as temporal features. Then, it exploits a
multi-regression model, including two types of SVR, to predict the final
score of videos.  A mixed neural network is derived in
\cite{you2019deep}. It uses a 3D convolutional neural network (3D-CNN)
and a Long-Short-Term Memory (LSTM) network as the feature extractor and
the quality predictor, respectively.  Following the VSFA method
\cite{li2019quality}, MDTVSFA \cite{li2021unified} adopts an unified
BVQA framework with a mixed-dataset training strategy to achieve better
prediction performance against authentic-distortion video datasets.  PVQ
\cite{ying2021patch} adopts a local-to-global region-based BVQA
architecture, which is trained with different kinds of patches. Also, a
large authentic-distortion video dataset is built and reported in
\cite{ying2021patch}.  QSA-VQM \cite{agarla2020no} uses two CNNs to
extract quality attributes and semantic content of each video frame,
respectively, and one Recurrent Neural Network (RNN) to estimate the
quality score of the whole video by incorporating temporal information.
To address the diverse range of natural temporal and spatial distortions
commonly observed in user-generated-content datasets, CNN-TLVQM
\cite{korhonen2020blind} integrates spatial features obtained from a CNN
model and handcrafted statistical temporal features obtained via TLVQM.
The CNN model was originally trained for image quality assessment using
a transfer learning technique. 

\begin{figure*}[!htbp]
\centering
\includegraphics[width=0.9\linewidth]{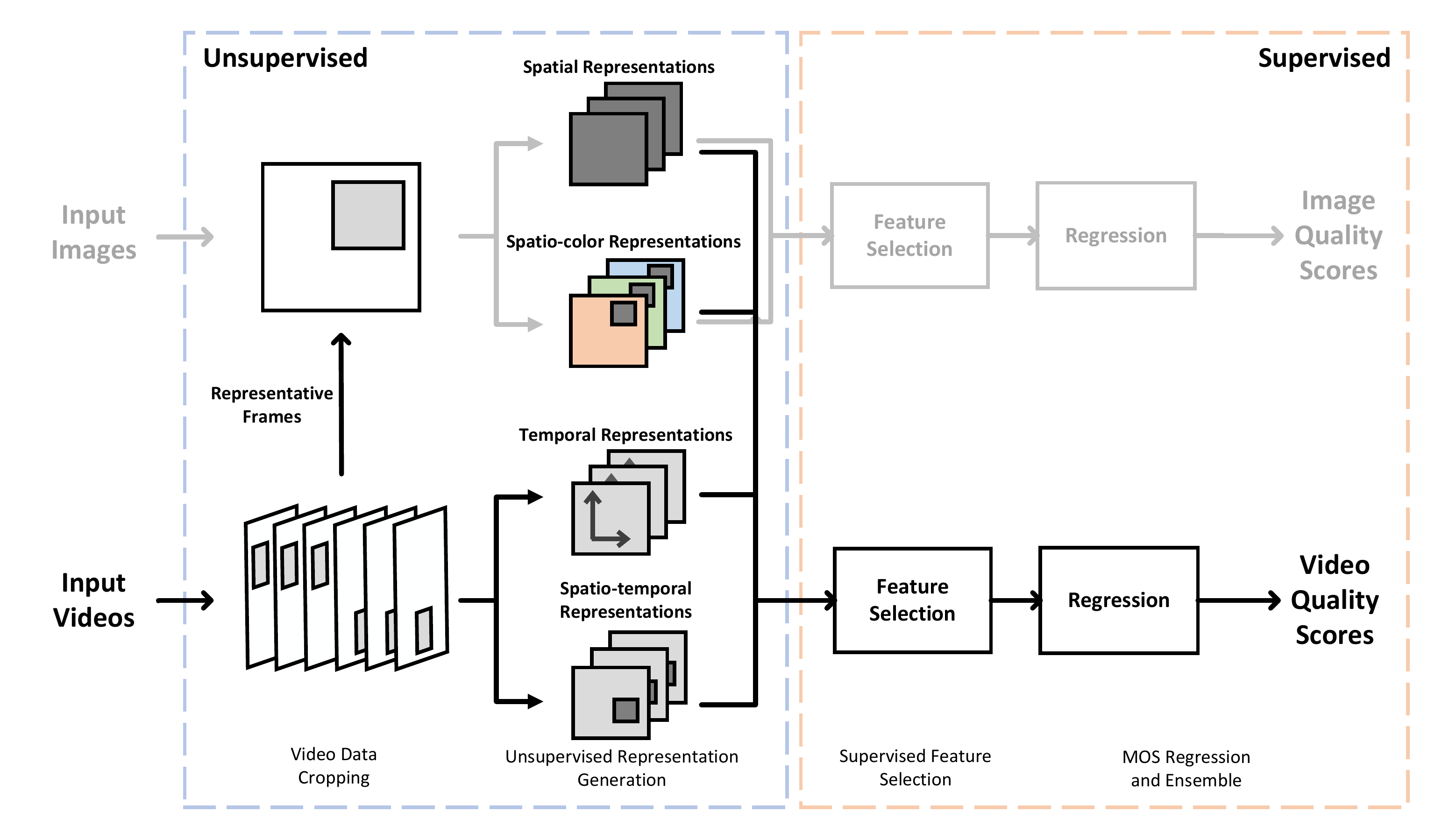}
\caption{The system diagram of the proposed GreenBVQA method.}\label{fig:pipeline}
\end{figure*}

\subsection{Video Quality Assessment at the Edge}\label{subsec:VEC}

Machine learning models have been extensively deployed on edge devices
\cite{qi2018enabling}.  Several video analytics tasks are implemented in
the edge computing platform \cite{xiao2021towards}. Besides prediction
accuracy, important metrics to be considered at the edge include
latency, computational complexity, memory size, etc.  The remarkable
success achieved by DL in various domains, such as computer vision and
natural language processing, has inspired the application of DL to edge
devices (say, smartphones and IoT sensors) \cite{chen2019deep}.  They
often generate a significant amount of data that demand local processing
due to the high data communication cost. Edge computing has emerged in
video analytics in recent years. One objective is to optimize the
tradeoff between accuracy and cost \cite{xiao2021towards}.  For example,
VideoEdge \cite{hung2018videoedge} is an edge-based video analysis tool
that is implemented in a distributed cloud-edge architecture, comprising
edge nodes and the cloud. 

Given heavy video traffics on mobile or edge networks, there is a
growing demand for higher transmission rates and lower network latency.
In order to tackle these challenges, adaptive bitrate (ABR) technologies
\cite{bentaleb2018survey, yoon2019hardware} are commonly used in video
distribution. The ABR scheme consists of two main components. First, the
video is encoded into multiple streaming versions with different bit
rates.  Second, each streaming version is segmented into multiple
segments based on the terminal capabilities and network conditions.
Consequently, the most suitable streaming version is dynamically
provided to the user.  The advantage of ABR lies in its ability to
reduce the occurrence of choppy videos while enhancing user's quality of
experience (QoE).

To enhance QoE at the edge, one essential technology is the automatic
measure of user's perceptual video quality.  As stated in
\cite{duanmu2020assessing}, a better understanding of human perceptual
experience and behavior is the most dominating factor in improving the
performance of ABR algorithms. Since reference videos are not available
on edge devices, BVQA methods become the only option.  State-of-the-art
BVQA methods rely on large pre-trained models and exhibit high
computational complexity, making them impractical for deployment on edge
devices.  Lightweight BVQA methods are in urgent need to address this
void. Along this line, Mirko \textit{et al.} \cite{agarla2021efficient}
propose an efficient BVQA method with two lightweight pre-trained
MobileNets \cite{howard2017mobilenets} with certain limitations, such as
degraded prediction accuracy.

\section{GreenBVQA Method}\label{sec:BVQA_method}

The system diagram of the proposed GreenBVQA method is shown in Fig.
\ref{fig:pipeline}.  It has a modularized system consisting of four
modules: 1) video data cropping, 2) unsupervised representation
generation, 3) supervised feature selection, and 4) MOS regression and
ensemble.  We will elaborate the operations in each module below. 

\begin{figure*}[!htbp]
\centering
\includegraphics[width=1.0\linewidth]{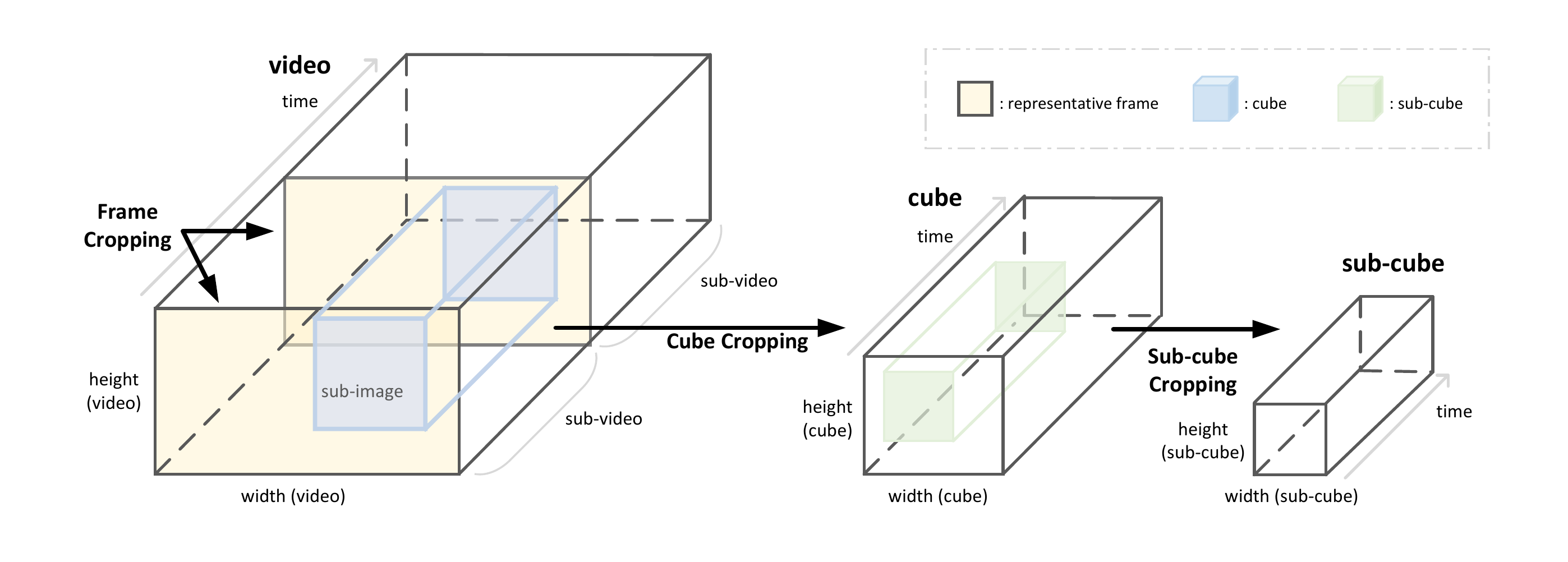}
\caption{The video data cropping module of GreenBVQA.}\label{fig:data_cropping}
\end{figure*}

\subsection{Video Data Cropping}

GreenBVQA adopts a hierarchical data cropping approach as illustrated in
Fig. \ref{fig:data_cropping}. This serves as a pre-processing step for
later modules.  Upon receiving an input video clip, we first split it
into multiple sub-videos in the time domain. For instance, a ten-second
video can be partitioned into ten non-overlapping sub-videos, each of
one-second duration.  The sub-video serves as the basic unit for future
processing.  Given a sub-video, we consider the following three cropping
schemes. 
\begin{enumerate}
\item Frame Cropping \\
One representative frame is selected from each sub-video. It can be the
first frame, an I-frame, or an arbitrary frame. Here, we aim to get
spatial-domain information. For example, several sub-images can be
cropped from a representative frame.  Both spatial and spatio-color
representations will be computed from each sub-image to be discussed in
the next subsection. 
\item Cube Cropping \\
We collect co-located sub-images from all frames in one sub-video as
shown in Fig. \ref{fig:data_cropping}.  This process is referred to as
``cube cropping" since it contains both spatial and temporal
information.  The purpose of cube cropping is to reduce the amount of
data to be processed in the later modules. This is needed as the data
size of videos is substantially larger than that of images. 
\item Sub-cube Cropping \\
We crop out a sub-cube from a cube that has a shorter length in the time
domain and a smaller size in the spatial domain (see Fig.
\ref{fig:data_cropping}). It is used to extract spatio-temporal
representations.  The rationale for sub-cube cropping is akin to that of
cube cropping - reducing the amount of data to be processed later. 
\end{enumerate} 
The rationale behind is that, considering the substantial video data
amount, it is impractical to process the entirety of the data.  Instead,
hierarchical data cropping is employed as a means to gradually curtail
the volume of data processing, while ensuring that excessive data is not
discarded abruptly at any one stage of the cropping process. 

\subsection{Unsupervised Representation Generation}

We consider the following four representations in GreenBVQA. 
\begin{enumerate}
\item Spatial representations. They are extracted from the Y channel of
sub-images cropped from representative frames. 
\item Spatio-Color representations. They are extracted from cubes of
size $(H \times W) \times C$, where $H$ and $W$ are the height and width
of sub-images and $C=3$ is the number of color channels, respectively. 
\item Temporal representations. They are the concatenation of
statistical temporal information of cubes. 
\item Spatio-Temporal representations. They are extracted from sub-cubes
of size $(H \times W) \times T$, where $H$ and $W$ are the height and
width of sub-images and $T$ is the length of sub-cubes in the time
domain, respectively. 
\end{enumerate}
GreenBVQA employs all four types of representations collectively to
predict perceptual video quality scores.  On the other hand, spatial and
spatio-color representations can be utilized to predict the quality scores
of individual images or sub-images. 

\begin{table*}[t]
\centering
\caption{The selected hyper-parameters for spatial representation
generation in our experiments, where the transform dimensions are
denoted by \{$(H \times W)\text{, }C$\} for spatial and channel
dimensions, the stride parameter is denoted as \{Spatial Stride$^2$\},
and L, M, and H represent low-frequency, mid-frequency, and high-frequency representations, respectively.}\label{table:s_arc}
\begin{tabular}{ l | c | c | c | c }
\hline\hline
Layer & \multicolumn{3}{c}{Spatial} \vline & Output Size\\\hline
Input &\multicolumn{3}{c}{$320 \times 320$} \vline &$320 \times 320$\\\hline
DCT & \multicolumn{3}{c}{$(8 \times 8) , 1$} \vline & $(40 \times 40) , 64$ \\\hline
Split & \multicolumn{2}{c}{Low-freq (L)} \vline  & High-freq (H) & 
\makecell{L: $(40 \times 40) , 1$ \\ H: $(40 \times 40) , 63$} \\\hline
Hop1 &\multicolumn{2}{c}{\makecell{{$(4 \times 4) , 1$} \\ {stride 2$^2$}}} \vline 
&- &\makecell{L: $(19 \times 19) , 16$ \\ H: $(40 \times 40) , 63$}\\\hline
Split & Low-freq (L)  & Mid-freq (M)  & High-freq (H) &\makecell{L: $(19 \times 19) , 3$ \\ 
M: $(19 \times 19) , 13$ \\ H: $(40 \times 40) , 63$} \\\hline
Pooling &{$(1 \times 1) , 3$}  &$(2 \times 2) , 13$  &$(4 \times 4) , 63$
&\makecell{L: $(19 \times 19) , 3$ \\ M: $(9 \times 9) , 13$ \\ 
H: $(10 \times 10) , 63$}\\\hline 
Hop2 &\makecell{{$(3 \times 3) , 3$} \\ {stride 2$^2$}} &- &-
&\makecell{L: $(9 \times 9) , 27$ \\ M: $(19 \times 19) , 13$ \\ 
H: $(10 \times 10) , 63$}\\\hline \hline
\end{tabular}
\end{table*}

\begin{table*}[t]
\centering
\caption{The selected hyper-parameters for
spatial-color representations generation, where the transform
dimensions are denoted by \{$(H \times W) \times T\text{, }C$\} for
spatial, color, and channel dimensions and $L$ and $H$
represent low-frequency and high-frequency representations,
respectively.}\label{table:sc_arc}
\begin{tabular}{ l | c | c | c || c | c | c }
\hline\hline
Layer & \multicolumn{2}{c}{Spatio-color} \vline & Output Size\\\hline
Input & \multicolumn{2}{c}{$(320 \times 320) \times 3$} \vline & 
$(320 \times 320) \times 3$ \\\hline
Pooling & \multicolumn{2}{c}{$(2 \times 2) \times 1$} \vline &$(160 
\times 160) \times 3$ \\\hline
Hop1 &\multicolumn{2}{c}{$(4 \times 4) \times 3$} \vline &$(40 \times 40) 
\times 1 , 48$ \\\hline
Split & Low-freq (L) & High-freq (H) & \makecell{L: $(40 \times 40) \times 1 , 
3$ \\ H: $(40 \times 40) \times 1 , 45$} \\\hline
Pooling &- &$(2 \times 2) \times 1 , 45$ &\makecell{L: $(40 \times 40) 
\times 1 , 3$ \\ H: $(20 \times 20) \times 1 , 45$} \\\hline
Hop2 &$(4 \times 4) \times 1 , 1$ &- & \makecell{L: $(10 \times 10) \times 1 , 48$ \\ 
H: $(20 \times 20) \times 1 , 45$} \\ \hline \hline
\end{tabular}
\end{table*}

\begin{table*}[t]
\centering
\caption{The selected hyper-parameters for 
spatial-temporal representations generation, where the transform
dimensions are denoted by \{$(H \times W) \times T\text{, }C$\} for
spatial, temporal, and channel dimensions and $L$ and $H$
represent low-frequency and high-frequency representations,
respectively.}\label{table:st_arc}
\begin{tabular}{ l |  c | c | c }
\hline\hline
Layer &  \multicolumn{2}{c}{Spatio-temporal} \vline &Output Size\\\hline
Input & \multicolumn{2}{c}{$(96 \times 96) 
\times 15$} \vline &$(96 \times 96) \times 15$\\\hline
Pooling & \multicolumn{2}{c}{-} \vline &$(96 \times 96) 
\times 15$\\\hline
Hop1  &\multicolumn{2}{c}{$(8 \times 8) \times 3$} \vline 
&$(12 \times 12) \times 5 , 192$\\\hline
Split  & Low-freq (L) & High-freq (H) & 
\makecell{L: $(12 \times 12) \times 5 , 4$ \\ H: $(12 \times 12) \times 5 , 188$} \\\hline
Pooling  &- &$(2 \times 2) \times 1 , 188$
&\makecell{L: $(12 \times 12) \times 5 , 4$ \\ H: $(6 \times 6) \times 5 , 188$}\\\hline
Hop2  &$(2 \times 2) \times 5 , 4$ &- & 
\makecell{L: $(6 \times 6) \times 1 , 80$ \\ H: $(6 \times 6) \times 5 , 188$}\\ \hline \hline
\end{tabular}
\end{table*}

\subsubsection{Spatial Representations}

\begin{figure}[!h]
\centering
\includegraphics[width=0.6\linewidth]{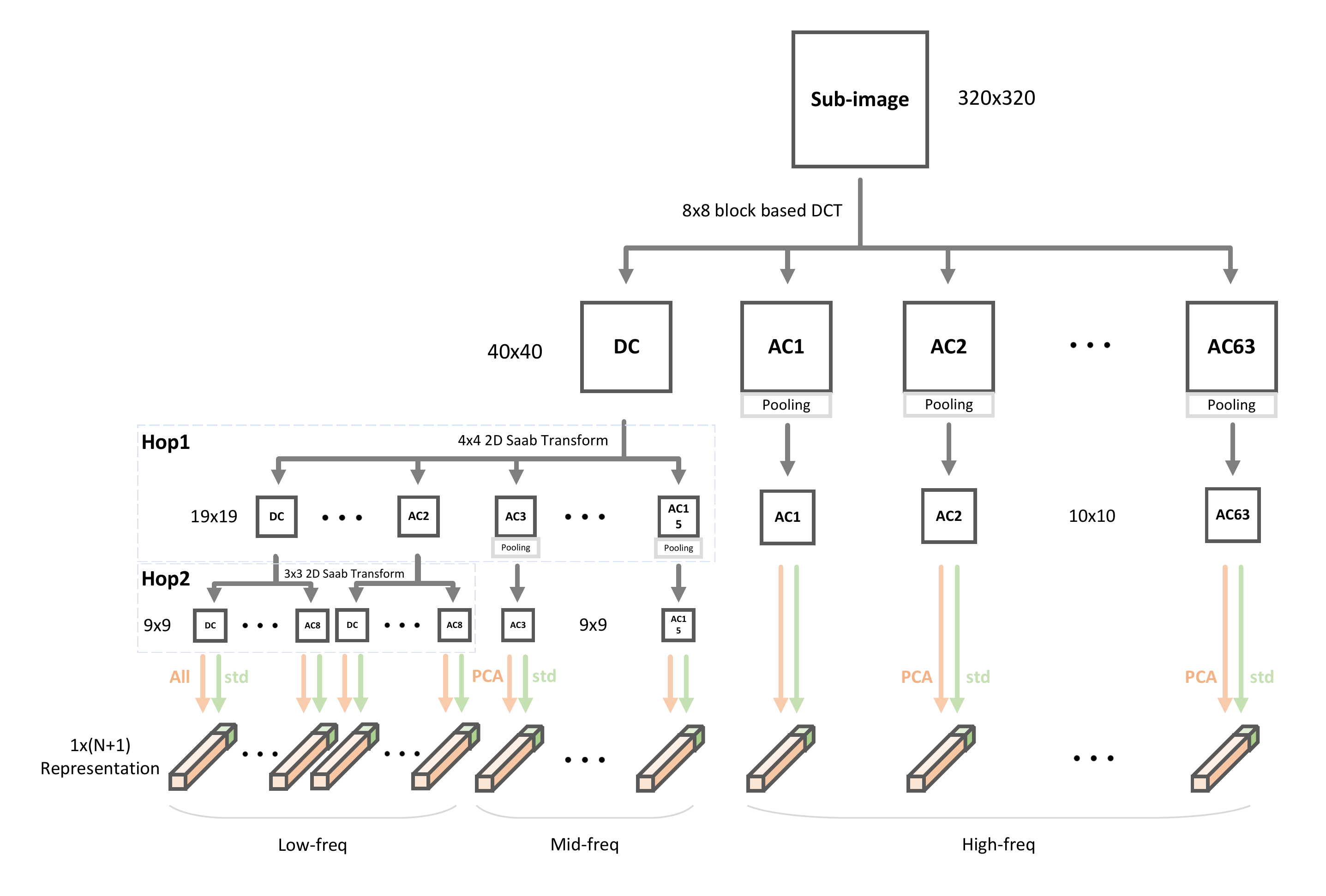}\\
\caption{The block diagram of unsupervised spatial representation 
generation.}\label{fig:structure_2d}
\end{figure}

As discussed in deriving the spatial representation for GreenBIQA
\cite{mei2023lightweight}, a three-layer structure is adopted to extract
local and global spatial representations from the sub-images. This is
summarized in Table \ref{table:s_arc} and depicted in Fig.
\ref{fig:structure_2d}.  Input sub-images are partitioned into
non-overlapping blocks of size $8 \times 8$, and the Discrete Cosine
Transform (DCT) coefficients are computed through the block DCT
transform.  These coefficients, consisting of one DC coefficient and 63
AC coefficients (AC1-AC63), are organized into 64 channels.  The DC
coefficients exhibit correlations among spatially adjacent blocks, which
are further processed by using the Saab transform
\cite{kuo2019interpretable}.  The Saab transform computes the patch
mean, referred to as the DC component, using a constant-element kernel.
Principal Component Analysis (PCA) is then applied to the mean-removed
patches to derive data-driven kernels, known as AC kernels.  The AC
kernels are applied to each patch, resulting in AC coefficients of the
Saab transform.  To decorrelate the DC coefficients, a two-stage
process, namely Hop1 and Hop2, is employed.  The coefficients obtained
from each channel, either with or without down-sampling at different
Hops and the DCT layer, are utilized to calculate standard deviations,
PCA coefficients, or are left unchanged.  According to the spectral
frequency in DCT and Saab domain, the coefficients from Hop2, Hop1, and the
DCT layer are denoted as low-frequency, mid-frequency, and
high-frequency representations, respectively.  Low- and mid-frequency
representations contain global information from large receptive fields,
while high-frequency representations contain information of details from
a small receptive field.  Then, all representations are concatenated to
form spatial representations. 

\begin{figure*}[t]
\centering
  \begin{tabular}{c @{\hspace{20pt}} c }
    \includegraphics[width=.45\linewidth]{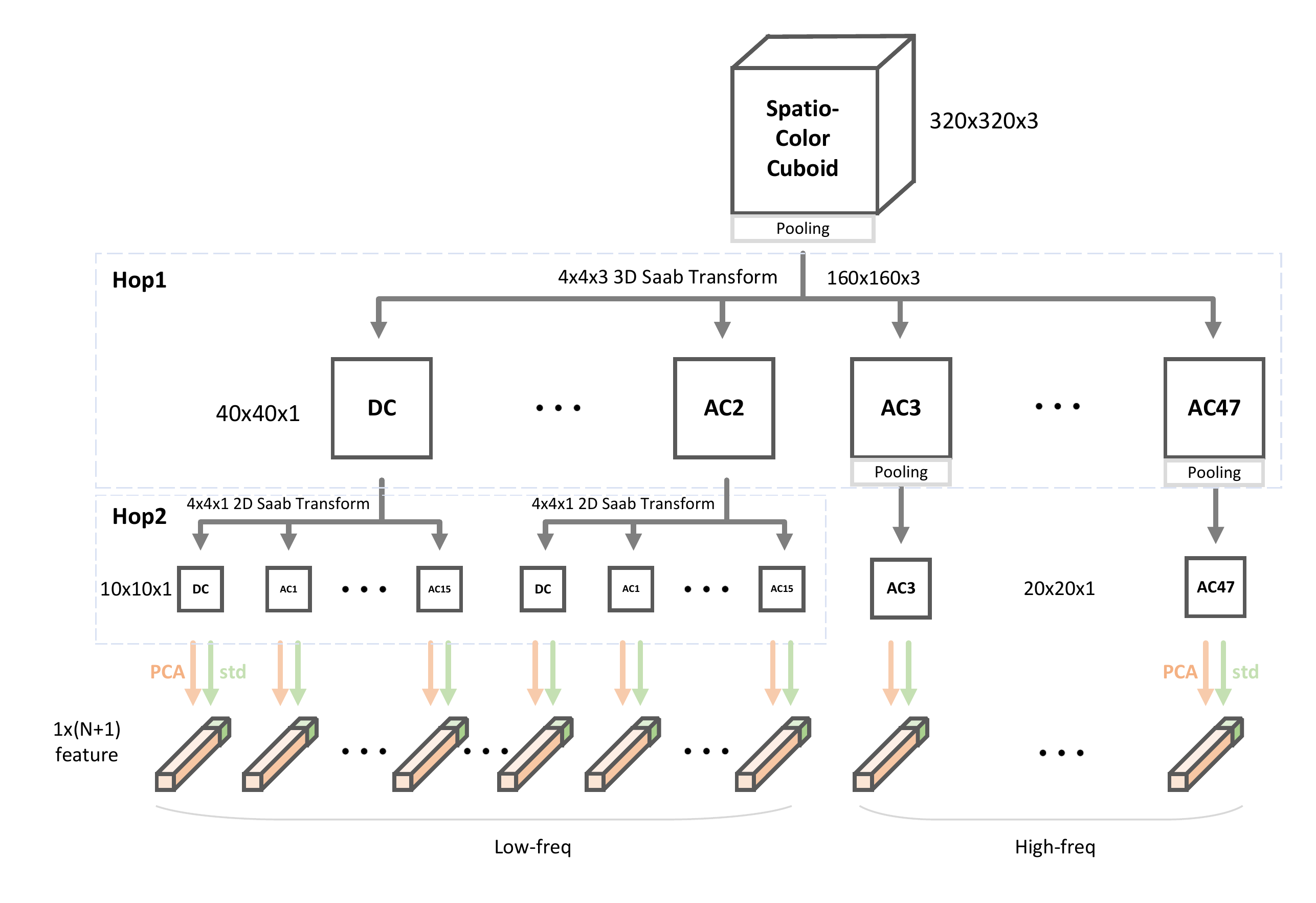} &
    \includegraphics[width=.45\linewidth]{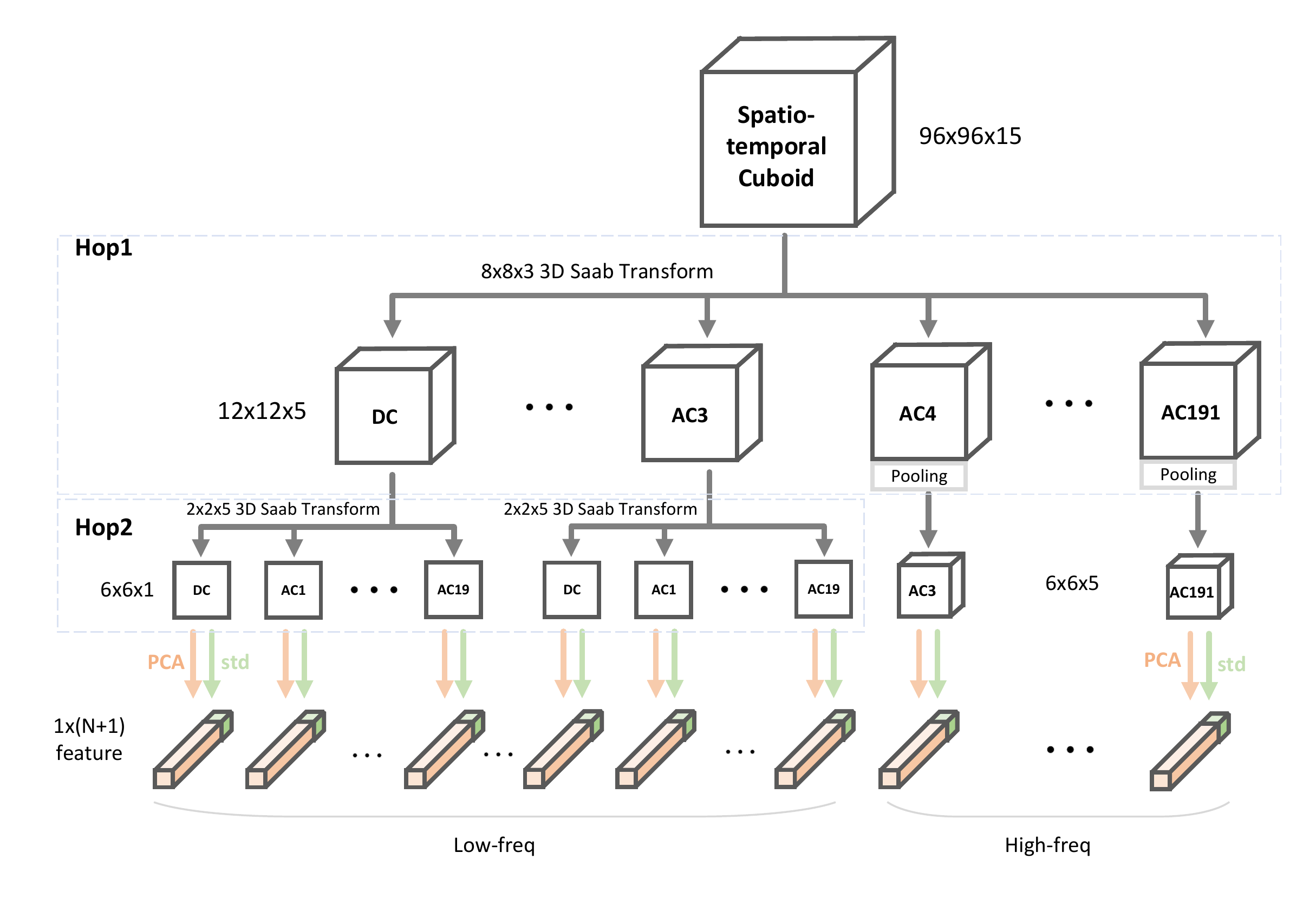}
    \\[\abovecaptionskip]
    \small (a) Generation of spatiao-color representations &
    \small (b) Generation of spatio-temporal representations
  \end{tabular}
\caption{The block diagram of unsupervised spatio-color and
spatio-temporal representations generation.}\label{fig:sc_st_generation}
\end{figure*}

\subsubsection{Spatio-Color Representations}

The representations for spatio-color cubes are derived using 3D Saab and
PCA methods. The hyper-parameters are given in Table
\ref{table:sc_arc}, and the data processing block diagram is depicted
in Fig. \ref{fig:sc_st_generation} (a).  A spatio-color cuboid has
dimensions of $H \times W \times C$, where $H$ and $W$ represent the
height and width of the sub-image respectively, and $C = 3$ denotes the
number of color channels. It is fed to a two-hop structure.  In Hop1,
it is divided into non-overlapping cuboids of size $4 \times
4 \times 3$, and the 3D Saab transform is applied individually,
resulting in one DC channel and 47 AC channels (AC1-AC47).  Each channel
has a spatial dimension of $40 \times 40$.  Since the DC, AC1, and AC2
coefficients exhibit high spatial correlation, in Hop2, a 2D Saab transform
is used to decompose these channels of size $40 \times 40$ into
non-overlapping blocks of size $4 \times 4$.  For the other 45 AC
coefficients obtained from Hop1, their absolute values are taken and a
$2 \times 2$ max pooling operation is performed, yielding 45 channels
with a spatial dimension of $20 \times 20$.  In total, we obtain 93
channels, comprising 48 low-frequency channels from Hop2 and 45
high-frequency channels from Hop1, with spatial size of $10 \times 10$
and $20 \times 20$, respectively.  The coefficients obtained from each
channel are
utilized to calculate standard deviations and PCA coefficients. These
computed coefficients are then concatenated to form spatio-color representations. 

\begin{table}[t]
\centering
\caption{Summary of 14-D raw temporal representations.} 
\label{table:temp_representation}
\begin{tabular}{ l | r }
\hline
Index &Computation Procedure \\
\hline
$f_1 - f_2$  &Compute the mean of x-mvs and y-mvs\\
$f_3 - f_4$  &Compute the standard deviation of x-mvs and y-mvs\\
$f_5 - f_6$  &Compute the ratio of significant x-mvs and y-mvs\\
$f_7 - f_8$  &Collect the maximum of x-mvs and y-mvs\\
$f_9 - f_{10}$  &Collect the minimum of x-mvs and y-mvs\\
$f_{11}$  &Compute the mean of magnitude of mvs\\
$f_{12}$  &Compute the standard deviation of magnitude of mvs\\
$f_{13}$  &Compute the ratio of significant magnitude of mvs\\
$f_{14}$  &Collect the maximum of magnitude of mvs\\ \hline
\end{tabular}
\end{table}

\subsubsection{Temporal Representation}

The spatial and spatio-color representations are extracted from the
sub-images on representative frames.  Both of them represent the
information within individual frames while disregarding the temporal
information across frames. Here, a temporal representation generation is
proposed to capture the temporal information from motion vectors (mvs). 

Consider a cube of dimensions $H \times W \times T$, where $H$ and $W$
represent the height and width of sub-images, respectively, and $T$ is
the number of frames in the time domain.  For each $H \times W$
sub-image within a cube, motion vectors of small blocks are computed (or
collected from compressed video streaming). They are denoted as ${\bf
V}=((x_1, y_1) \cdots, (x_n, y_n))^T$, where $(x_n, y_n)$ represents the
motion vector of the $n^{th}$ block.  Specifically, $x_n$ and $y_n$ are the
horizontal magnitude and vertical magnitude of the motion vector and are named
x-mv and y-mv, respectively.  The magnitude of the motion vector can be
computed by $\sqrt{x_n^2 + y_n^2}$. 

The motion representation of a cube is computed based on its motion
vectors. This statistical analysis yields a 14-D temporal representation
of each sub-image as shown in Table \ref{table:temp_representation}.
They are arranged in chronological order to form raw temporal
representations.  Furthermore, PCA is applied to them to derive spectral
temporal representations.  Finally, the raw and spectral temporal
representations are concatenated to form the final temporal
representations. 

\subsubsection{Spatio-Temporal Representation}

Both spatial representations from sub-images of representative frames
and temporal representations from cubes are extracted individually from
a single domain.  It is also important to consider the correlation
between spatial information and temporal information in subjective score
prediction, as subjective assessments often take both aspects into
account when providing scores.  To extract spatio-temporal features from
both spatial and temporal domains, a two-hop architecture is adopted,
where the 3D Saab transform is conducted as depicted in Fig.
\ref{fig:sc_st_generation} (b).  The hyper-parameters are summarized in
Table \ref{table:st_arc}

The dimension of the spatio-temporal cube, which is the same as
sub-cubes in Fig.  \ref{fig:data_cropping}, is $H \times W \times T$,
where $H$, $W$ and $T$ represent the height, width, and the frame number
of the sub-cube, respectively.  These sub-cubes are fed into a two-hop
architecture, where the first and the second hops are used to capture
local and global representations, respectively.  The procedure used to
generate spatio-temporal representation is similar to that for the
spatio-color representation generation, except the 3D channel-wise Saab
transform is applied in both hops.  In Hop 1, we split the input
sub-cubes into non-overlapping 3-D cuboids of size $8 \times 8 \times
3$. They are converted to one-dimensional vectors for Saab coefficient
computation, leading to one DC channel and 191 AC channels, denoted by
AC1-AC191.  The size of each channel is $12 \times 12 \times 5$.  The
coefficients in DC and low-frequency AC (e.g., AC1-AC3) channels are
spatially and temporally correlated because the adjacent $8 \times 8
\times 3$ cuboids are strongly correlated.  Therefore, another 3-D Saab
transform is applied in Hop 2 to decorrelate the DC and low-frequency AC
channels from Hop 1. Similarly, we split these channels into several
non-overlapping 3D cuboids of size $2 \times 2 \times 5$.  Coefficients
in each cuboid are flattened into a 20-D vector denoted by ${\bf
y}=(y_1, \cdots, y_{20})^T$, and their Saab coefficients are computed.
The DC coefficients in Hop 2 are computed by the mean of the 20-D vector,
$\bar{y}=(\sum_{i=1}^{20} y_i)/20$. The remaining 19 AC coefficients,
denoted by AC1 to AC19, are generated by the principle component
analysis (PCA) on the mean-removed 20-D vector. 

To lower the number of coefficients that need to be processed, blocks in Hop
1 are downsampled to cuboids of size $6 \times 6 \times 5$ by using $2
\times 2$ max pooling in the spatial domain.  Given low-frequency
channels of size $6 \times 6 \times 1$ from Hop 2 and downsampled $6
\times 6 \times 5$ high-frequency channels from Hop 1, we generate 
two sets of representations as follows.
\begin{itemize}
\item The coefficients in each channel are first flattened to 1-D
vectors. Next, we conduct PCA and select the first $N$ PCA coefficients
of each channel to form the spectral features. 
\item We compute the standard deviation of coefficients from the 
same channel across the spatio-temporal domain. 
\end{itemize}
Finally, we concatenate the two sets of representations to form
the spatio-temporal representations.

\subsection{Supervised Feature Selection}

The number of unsupervised representations is large.  To reduce the
dimension of unsupervised representations, we select quality-relevant
features from the 4 sets of representations obtained in the unsupervised
representation generation part, by adopting the relevant feature test (RFT)
\cite{yang2022supervised}.  RFT enables the calculation of independent
losses for each representation, with lower loss values indicating
superior representations.  The RFT procedure involves splitting the
dynamic range of a representation into two sub-intervals using a set of
partition points.  For a given partition, the means of the training
samples in the left and right regions are computed as representative
values, and their respective mean-squared errors (MSE) are calculated.
By combining the MSE values of both regions, a weighted MSE for the
partition is obtained.  The search for the minimum weighted MSE across
the set of partition points determines the cost function for the
representation. It is important to note that RFT is a supervised feature
selection algorithm as it utilizes the labels of the training samples.

\begin{figure}[!htbp]
\centering
\includegraphics[width=0.6\linewidth]{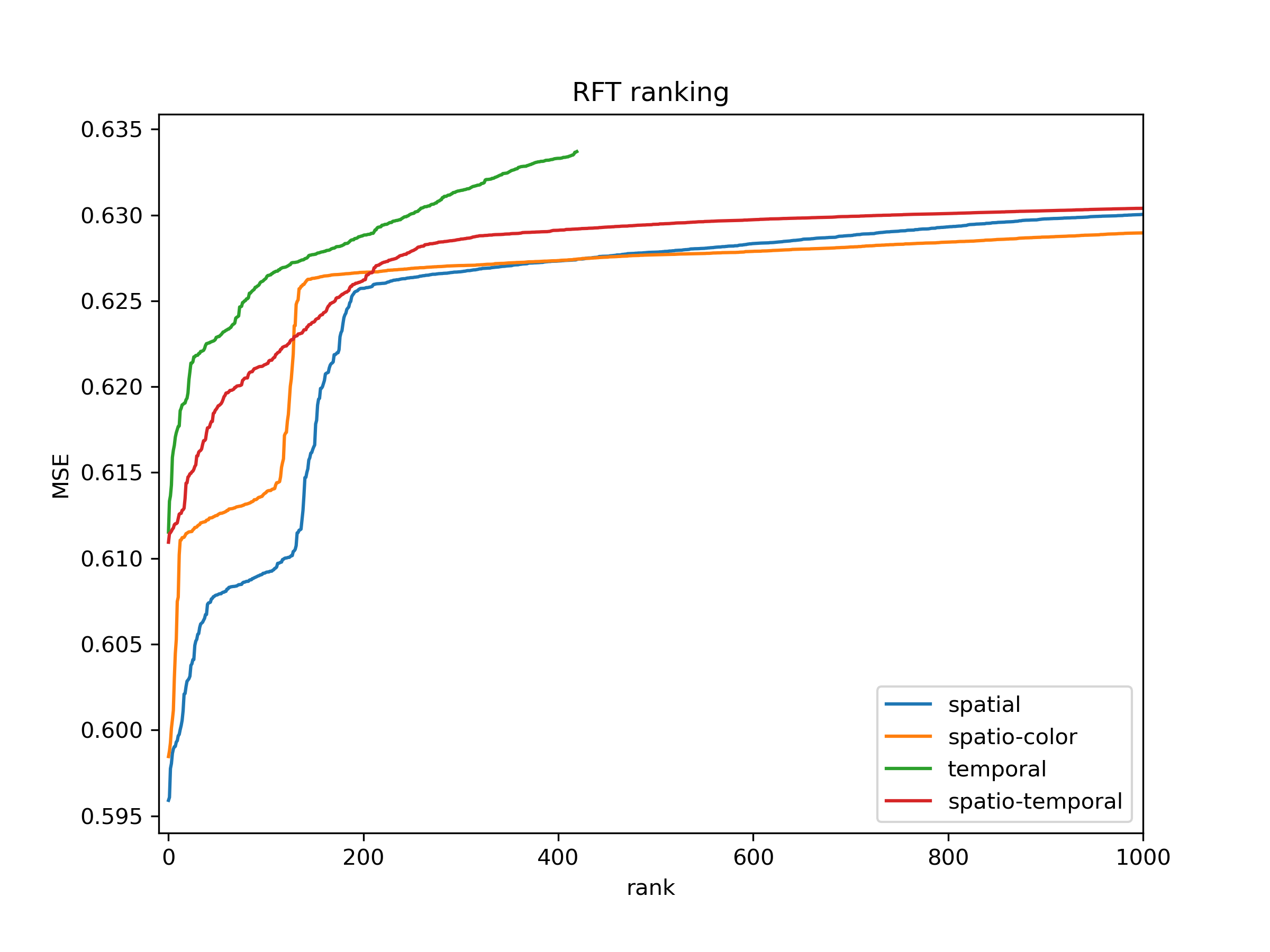}
\caption{RFT results of spatial, spatio-color, temporal, 
and spatio-temporal representations.}\label{fig:ranking}
\end{figure}

We computed the RFT results for spatial, spatio-color, temporal, and
spatio-temproal representations individually.  Fig. \ref{fig:ranking}
illustrates the sorting of representation indices based on their MSE
values, with separate curves.  The order of representations based on
their MSE values implies that dimensions of representation with lower
MSE are more likely to possess discriminative characteristics, thus
signifying their potential value as features. In order to effectively
collect features with discriminative attributes while concurrently
reducing the dimensionality of the representations, an efficient
strategy involves the selection of the top-ranked representation indices
derived from the RFT results.  The identification of relevant
representation subsets is facilitated by pinpointing positions near the
elbow points on each curve.  Consequently, considering the existence of
four distinct types of unsupervised representations, a subset comprising
the highest-ranked indices for each representation type is chosen.  The
selected indices are then concatenated to constitute a set of supervised
quality-relevant features for each cube. The dimensions of the
unsupervised representations and the supervised features chosen in this
manner are shown in Table \ref{table:representation_feature}, with
particular focus on their application to the KoNViD-1k dataset as
delineated in \cite{hosu2017konstanz}. 

\begin{table}[t]
\centering
\caption{Dimensions of unsupervised representation 
and selected supervised features on KoNViD-1k dataset.}
\label{table:representation_feature}
\begin{tabular}{ l | c c } \hline
 &Representation &Feature \\
&Dimension &Dimension \\
\hline
Spatio  &6,637 &220\\
Spatio-color &6,793 &200\\
Temporal &420 &140\\
Spatio-temporal &8,878 &240\\
\hline
Sum &22,728 &800\\
\hline
\end{tabular}
\end{table}

\subsection{MOS Regression and Ensembles}

Once the quality-relevant features are selected, we employ the XGBoost
\cite{chen2016xgboost} regressor as the quality score prediction model
that maps $d$-dimensional quality-relevant features to a single quality
score.  After the regressor's prediction, each cube is assigned a predicted
score.  The scores of cubes belonging to the same sub-video are then
ensembled using a median filter, resulting in the score of the
sub-video, which predict the Mean Opinion Score (MOS) of a short interval of
frames from the input video.  To obtain the final MOS for the entire
input video, a mean filter is applied to aggregate the scores from all
sub-videos belonging to the same input video. 

\begin{table*}[t]
\centering
\caption{Statistics of three VQA datasets.}\label{table:dataset_BVQA}
\begin{tabular}{ l | cccccc }
\hline
Dataset & Ref. &Scenes. &Resolution &Time Duration &MOS range &Environment\\
\hline
CVD2014 \cite{nuutinen2016cvd2014}      &234 &5 &480p, 720p &10-25s &[-6.50, 93.38] &In-lab\\
KoNViD-1k \cite{hosu2017konstanz} &1,200 &1,200 &540p &8s &[1.22, 4.64] &Crowdsourced\\
LIVE-VQC \cite{sinno2018large}   &585 &585 &240p-1080p &10s &[6.2237, 94.2865] &Crowdsourced\\
\hline
\end{tabular}
\end{table*}

\section{Experiments}\label{sec:experiments}

\subsection{Experiments Setup}

We discuss VQA datasets, performance benchmarking methods, evaluation
metrics, and some implementation details below. 

\subsubsection{Datasets}

We evaluate GreenBVQA on three VQA datasets: CVD2014
\cite{nuutinen2016cvd2014}, KoNViD-1k \cite{hosu2017konstanz}, and
LIVE-VQC \cite{sinno2018large}. Their statistics are summarized in Table
\ref{table:dataset_BVQA}.  CVD2014 is captured in a controlled
laboratory environment. Thus, it is also called the lab-generated
dataset.  It comprises 234 video sequences of resolution $640 \times
480$ or $1280 \times 720$. They are acquired with 78 cameras ranging
from low-quality mobile phones to high-quality digital single-lens
reflex cameras.  Each video displays one of five scenes with distortions
associated with the video acquisition process.  KoNViD-1k and LIVE-VQC
are authentic-distortion datasets, also known as
user-generated content (UGC) datasets.  KoNViD-1k comprises 1200 video
sequences, each of which lasts for 8 seconds with a fixed resolution.
LIVE-VQC consists of a collection of video sequences of a fixed duration
in multiple resolutions.  Both of them contain diverse content and a
wide range of distortions. 

\subsubsection{Benchmarking Methods}

We compare the performance of GreenBVQA with eleven benchmarking methods
in Table \ref{table:BVQA_individual}. These methods can be classified
into three categories. 
\begin{itemize}
\item Three conventional BIQA methods: NIQE \cite{mittal2012making},
BRISQUE \cite{mittal2012no}, and CORNIA \cite{ye2012unsupervised}.  They
are applied to frames of distorted videos. Then, the predicted
scores are ensembled to yield the ultimate BVQA score. 
\item Three conventional BVQA methods without neural networks: V-BLIINDS
\cite{saad2014blind}, TLVQM \cite{korhonen2019two}, and VIDEVAL
\cite{tu2021ugc}. 
\item Five state-of-the-art DL-based methods with pre-trained models:
VSFA \cite{li2019quality}, RAPIQUE \cite{tu2021rapique}, QSA-VQM
\cite{agarla2020no}, Mirko \textit{et al.} \cite{agarla2021efficient},
and CNN-TLVQM \cite{korhonen2020blind}. They are also called advanced DL methods.
\end{itemize}

\subsubsection{Evaluation Metrics}

The MOS prediction performance is measured by two well-known metrics:
the Pearson Linear Correlation Coefficient (PLCC) and the Spearman Rank
Order Correlation Coefficient (SROCC).  PLCC is employed to assess the
linear correlation between the predicted scores and the subjective
quality scores.  It is defined as
\begin{equation}
\textit{PLCC} = \frac{\sum_{i}(p_i - p_m)(\hat{p_i}-\hat{p_m})}
{\sqrt{\sum_{i}(p_i - p_m)^2}\sqrt{\sum_{i}(\hat{p_i}-\hat{p_m})^2}},
\end{equation}
where $p_i$ and $\hat{p_i}$ denote the predicted score and the
corresponding subjective quality score, respectively, for a test video
sample. Additionally, $p_m$ and $\hat{p_m}$ represent the means of the
predicted scores and subjective quality scores, respectively.  SROCC
is used to measure the monotonic relationship between the predicted
scores and the subjective quality scores, considering the relative
ranking of the samples.  It is defined as
\begin{equation}
\textit{SROCC} = 1 -  \frac{6\sum_{i=1}^{L}(m_i - n_i)^2} {L(L^2-1)},
\end{equation}
where $m_i$ and $n_i$ represent the ranks of the predicted score $p_i$
and the corresponding subjective quality score $\hat{p_i}$,
respectively, within their respective sets of scores. The variable $L$
represents the total sample number.

\subsubsection{Implementation Details}

{\bf Video Data Cropping.} Each sub-video has a length of 30 frames.  To
derive a representative frame for each of these sub-videos, a
straightforward selection process designates the first frame of each
sub-video.  From each such representative frame, a set of six
sub-images, each possessing dimensions of $320 \times 320$ pixels, is
randomly cropped.  Consequently, each cube consists of $(320 \times 320)
\times 30$ pixels, and a single sub-video harboring a complement of six
such cubes.  Specifically, in the context of generating spatio-temporal
representations, sub-cubes are derived from these cubes.  These
sub-cubes are intentionally configured to assume dimensions of $(96
\times 96) \times 15$.  The dimensions $(96 \times 96)$ denote the
central region cropped from the sub-images, while the sub-cubes'
temporal component is constructed by collecting frames at every
two-frame intervals within their corresponding cubes.  Notably, the
selection process ensures that only a solitary sub-cube is cropped from
each cube. 

{\bf Unsupervised Representation Generation.} For spatial representation
generation, the $8 \times 8$ DCT transform and the $4 \times 4$ Saab
transform are used to generate spatial representations of 6,637
dimensions, as shown in Table \ref{table:s_arc}.  Similarly, in
accordance with the structure outlined in Table \ref{table:sc_arc} and Table \ref{table:st_arc},
the dimensions of spatio-temporal and spatio-color representations are
8,878 and 6,793, respectively.  Furthermore, it is noteworthy that the
generation of temporal representations involves the computation of
14-dimensional representations, as shown in Table
\ref{table:temp_representation}, for each individual frame within a
cube. Consequently, the cumulative temporal representation for each cube
aggregates to a dimensionality of 420. 

{\bf Supervised Feature Selection.} Following the application of RFT to
each type of representation generated from the KoNViD-1k dataset,
independently, the resulting selected features exhibit dimensions of 220
for spatial features, 200 for spatio-color features, 140 for temporal
features, and 240 for spatio-temporal features.  It is important to note
that the dimensions of the selected features may vary across different
datasets, as the distribution of data and content can differ among
various datasets. 

{\bf MOS Regression and Ensembles.} The XGBoost regressor is used to
train and predict the MOS score of each cube. The max depth of each tree
is 5 and the subsampling rate is 0.6. The maximum number of trees is
2,000 with early termination. Given the score of each cube, a median
filter is used to obtain the score of each sub-video. Next, we take the
average of all sub-videos' scores to obtain the final score of the input
video. 

{\bf Performance Evaluation.} To ensure reliable evaluation, we
partition a VQA dataset into two disjoint sets: the training set (80\%)
and the testing set (20\%). We set 10\% aside in the training set for
validation purpose.  We conduct experiments in 10 runs and report
the median values of PLCC and SROCC. 

\begin{table*}[!htbp]
\centering
\caption{Comparison of the PLCC and SROCC performance of 10 benchmarking
methods against three VQA datasets.}\label{table:BVQA_individual}
\begin{tabular}{l c|c c c c c c c c}
\hline
& & \multicolumn{2}{c}{CVD2014}
& \multicolumn{2}{c}{LIVE-VQC} & \multicolumn{2}{c}{KoNViD-1k} & \multicolumn{2}{c}{Average}\\\hline
&Model & SROCC$\uparrow$ & PLCC$\uparrow$ & SROCC$\uparrow$ & PLCC$\uparrow$ & SROCC$\uparrow$ & PLCC$\uparrow$ & SROCC$\uparrow$ & PLCC$\uparrow$\\\hline
&NIQE\cite{mittal2012making} &0.475 &0.607 &0.593 &0.631 &0.539 &0.551 &0.535 &0.596\\
&BRISQUE\cite{mittal2012no} &0.790 &0.804 &0.593 &0.624 &0.649 &0.651 &0.677 &0.654\\ 
&CORNIA\cite{ye2012unsupervised} &0.627 &0.663 &0.681 &0.723 &0.735 &0.735 &0.681 &0.707\\
\hline
&V-BLIINDS\cite{saad2014blind} &0.795 &0.806 &0.681 &0.699 &0.706 &0.701 &0.727 &0.735\\
&TLVQM\cite{korhonen2019two} &0.802 &0.823 &0.783 &0.785 &0.763 &0.765 &0.782 &0.791\\
&VIDEVAL\cite{tu2021ugc} &0.814 &0.832 &0.744 &0.748 &0.770 &0.771 &0.776 &0.783\\
\hline
&VSFA\cite{li2019quality} &0.850 &\underline{0.859} &0.717 &0.770 &0.794 &0.798 &0.787 &0.809\\
&RAPIQUE\cite{tu2021rapique} &0.807 &0.823 &0.741 &0.761 &0.788 &\underline{0.805} &0.778 &0.796\\
&QSA-VQM \cite{agarla2020no} &\underline{0.850} &\underline{0.859} &0.742 &0.778 &\underline{0.801} &0.802 &0.797 &\underline{0.813}\\
&Mirko \textit{et al.}\cite{agarla2021efficient} &0.834 &0.848 &0.742 &0.780 &0.772 &0.784 &0.782 &0.804\\
&CNN-TLVQM \cite{korhonen2020blind} &\textbf{0.852} &\textbf{0.868} &\textbf{0.811} &\textbf{0.828} &\textbf{0.814} &\textbf{0.817} &\textbf{0.825} &\textbf{0.837}\\
 \hline
&GreenBVQA(Ours) &0.835 &0.854 &\underline{0.785} &\underline{0.789} &0.776 &0.779 &\underline{0.798} &0.807
\\\hline
\end{tabular}
\end{table*}

\subsection{Performance Comparison}

\subsubsection{Same-Domain Training Scenario}

We compare the PLCC and SROCC performance of GreenBVQA with that of the
other eleven benchmarking methods in Table \ref{table:BVQA_individual}.
GreenBVQA outperforms all three conventional BIQA methods (i.e., NIQE,
BRISQUE, and CORNIA) and all three conventional BVQA methods (i.e.,
V-BLIINDS, TLVVQM, and VIDEVAL) by a substantial margin in all three
datasets.  This shows the effectiveness of GreenBVQA in extracting
quality-relevant features to cover diverse distortions and content
variations.  GreenBVQA is also competitive with the five DL-based
BVQA methods.  Specifically, GreenBVQA achieves the second-best
performance for the LIVE-VQC dataset. It also ranks second in the
average performance of SROCC across all three datasets. 

As to the five DL-based BVQA methods, the performance of GreenBVQA is
comparable with that of QSA-VQM.  However, there exists a performance
gap between GreenBVQA and CNN-TLVQM, which is a state-of-the-art
DL-based method employing pre-trained models.  The VQA datasets,
particularly user-generated content datasets, pose significant
challenges due to non-uniform distortions across videos and a wide
variety of content without duplication.  Pre-trained models, trained on
large external datasets, have an advantage in extracting features for
non-uniform distortions and unseen content.  Nonetheless, these advanced
DL-based methods come with significantly larger model sizes and
inference complexity as analyzed in Sec. \ref{subsec:model_complexity}. 

\subsubsection{Cross-Domain Training Scenario}

To evaluate the generalizability of BVQA methods, we investigate the setting
where training and testing data come from different datasets.  Here, we
focus on the two UGC datasets (i.e., KoNViD-1k and LIVE-VQC) due to
their practical significance.  Two settings are considered: I) trained
with KoNViD-1k and tested on LIVE-VQC, and II) trained with LIVE-VQC and
tested on KoNViD-1k.  We compare the SROCC performance of GreenBVQA and
five benchmarking methods under these two settings in Table
\ref{table:BVQA_cross}.  The five benchmarking methods include two
conventional BVQA methods (TLVQM, and VIDEVAL) and three DL-base BVQA
methods (VSFA, QSA-VQM, and Mirko \textit{et al.}). 

We see a clear performance drop for all methods in the cross-domain
condition by comparing Tables \ref{table:BVQA_individual} and
\ref{table:BVQA_cross}. We argue that setting II provides a more
suitable scenario to demonstrate the robustness (or generalizability) of
a learning model.  This is because KoNViD-1k has a larger video number
and scene number as shown in Table \ref{table:dataset_BVQA}. Thus, we
compare the performance gaps in Table \ref{table:BVQA_cross} under
Setting II with those in the KoNViD-1k/SROCC column in Table
\ref{table:BVQA_individual}.  The gaps between VSFA, QSA-VQM, and CNN-TLVQM
against GreenBVQA become narrower for KoNViD-1k. They are down from
0.019, 0.023, and 0.038 (trained by the same dataset) to 0.015, -0.066,
and 0.024 (trained by LIVE-VOC), respectively.  
This suggests a high potential for GreenBVQA in the cross-domain training setting. 

\begin{table}[!htbp]
\centering
\caption{Comparison of the SROCC performance under the cross-domain
training scenario.}\label{table:BVQA_cross}
\begin{tabular}{l c| c | c}
\hline
& Settings  & I  & II \\\hline
& Training &KoNViD-1k &LIVE-VQC\\\hline
& Testing &LIVE-VQC &KoNViD-1k\\\hline
&TLVQM\cite{korhonen2019two} &0.572 &0.639\\
&VIDEVAL\cite{tu2021ugc} &0.591 &0.656\\
&VSFA\cite{li2019quality} &0.593 &0.671 \\
&QSA-VQM \cite{agarla2020no} &0.660 &0.590\\
&CNN-TLVQM \cite{korhonen2020blind} &\textbf{0.720} &\textbf{0.680}\\ \hline
&GreenBVQA(Ours) &0.631 &0.656 \\\hline
\end{tabular}
\end{table}

\begin{table*}[!htbp]
\centering
\caption{Model complexity comparison, where the reported SROCC and PLCC 
performance numbers are against the KoNViD-1k dataset.}\label{table:flop}
\begin{tabular}{ l c | c c c c c c c} \hline
&Model & SROCC$\uparrow$ & PLCC$\uparrow$ & Model Size (MB)$\downarrow$ &FLOPs$\downarrow$\\\hline
&VSFA\cite{li2019quality} &0.794 &0.798 &100.2 (15.8$\times$) &20T (1250$\times$)\\
&QSA-VQM \cite{agarla2020no} &0.801 &0.802 &196 (30.8$\times$) &40T (2500$\times$)\\
&Mirko \textit{et al.}\cite{agarla2021efficient} &0.772 &0.784 &42.3 (6.6$\times$) &1.5T (94$\times$)\\
&CNN-TLVQM \cite{korhonen2020blind} &\textbf{0.814} &\textbf{0.817} &98 (15.4$\times$) &21T (1312$\times$)\\ \hline
&GreenBVQA(Ours) &0.776 &0.779 &\textbf{6.36 (1$\times$)} &\textbf{16G (1$\times$)}\\ \hline
\end{tabular}
\end{table*}

\subsection{Comparison of Model Complexity}\label{subsec:model_complexity}

We evaluate the model complexity of various BVQA methods in three
aspects: model size, inference time, and computational complexity. 

\subsubsection{Model sizes}

There are two ways to measure the size of a learning model: 1) the
number of model parameters and 2) the actual memory usage.
Floating-point and integer model parameters are typically represented by
4 bytes and 2 bytes, respectively.  Since a great majority of model
parameters are in the floating point format, the actual memory usage is
roughly equal to $4 \times \mbox{(no. of model parameters)}$ bytes.
Here, we use the ``model size" to refer to actual memory usage below.
The model sizes of GreenBVQA and four benchmarking methods are compared
in Table \ref{table:flop}.  The size of the GreenBVQA model includes:
the representation generator (4.28MB) and a regressor (2.08MB),
leading to a total of 6.36 MB.  As compared with four DL-based
benchmarking methods, GreenBVQA achieves comparable SROCC and PLCC
performance with a much smaller model size.  

\begin{table*}[!htbp]
\centering
\caption{Inference time comparison in seconds for three videos selected from the considered databases. \{xxx\}frs@\{yyy\}p indicates the video frame length and the resolution, respectively.}\label{table:inference_time}
\begin{tabular}{l | c | c | c | c  }\hline
Mode &Model & \textbf{240frs@540p} & \textbf{364frs@480p} & \textbf{467frs@720p}\\\hline
\multirow{ 8 }{*}{CPU}
&V-BLIINDS\cite{saad2014blind} &382.06 &361.39 &1391.00 \\
&QSA-VQM \cite{agarla2020no} &281.21 &256.13 &900.72 \\
&VSFA\cite{li2019quality} &269.84 &249.21 &936.84 \\
&TLVQM\cite{korhonen2019two} &50.73 &46.32 &136.89 \\
&NIQE\cite{mittal2012making} &45.65 &41.97 &155.90 \\
&BRISQUE\cite{mittal2012no} &12.69 &12.34 &41.22\\ 
&Mirko \textit{et al.}\cite{agarla2021efficient} &8.43 &6.24 &16.29\\
&GreenBVQA(ours) &\textbf{3.22} &\textbf{4.88} &\textbf{6.26} \\\hline
\multirow{ 4 }{*}{GPU}
&QSA-VQM \cite{agarla2020no} &9.7 &9.15 &25.79 \\
&VSFA\cite{li2019quality} &8.85&7.55 &27.63 \\
&Mirko \textit{et al.}\cite{agarla2021efficient} &0.69 &0.85 &1.71\\
&GreenBVQA(ours) &\textbf{0.52} &\textbf{0.84} &\textbf{1.31} \\\hline
\end{tabular}
\end{table*}

\subsubsection{Inference time}

One measure of computational efficiency is inference time of predicting
video quality scores.  We compare the inference time of various BVQA
methods on a desktop with an Intel Core i7-7700 CPU@3.60GHz, 16 GB DDR4
RAM 2400 MHz, and a NVIDIA Titan X Pascal with 3840 CUDA cores. The
benchmarking methods include NIQE, BRISQUE, TLVQM, Mirko \textit{et
al.}, V-BLIINDS, VSFA, and QSA-VQM.  As shown in Table
\ref{table:inference_time}, we conduct experiments on three test videos
of various lengths and resolutions: a 240-frame video of resolution of
$960\times540$, a 346-frame video of resolution of $640\times480$, and a
467-frame video of resolution $1280\times720$.  We repeat the test for
each method ten times and report the average inference time (in seconds)
in Table \ref{table:inference_time}.  In the CPU mode, GreenBVQA has a
significantly shorter inference time as compared to other methods across
all resolutions. The efficiency gap widens as the video resolution
increases.  It is approximately 2.1x faster than Mirko \textit{et al.}
on average, which is the second most efficient method.  Furthermore,
GreenBVQA provides comparable performance with Mirko \textit{et al.} in
prediction accuracy as shown in Table \ref{table:BVQA_individual}, while
demanding a smaller model size. GreenBVQA can process videos in real
time, achieving an approximate speed of 75 frames per second, solely
relying on a CPU. 

It is worthwhile to mention that, as an emerging trend, edge computing
devices will contain heterogeneous computing units such as CPUs, GPUs,
and APUs (AI processing units). In light of the hardware diversity, our
evaluation extends to include a comparative analysis of GreenBVQA with
three DL-based methods that exploit GPU acceleration.  Leveraging mature
coding libraries and environments adept at facilitating deep learning
computations, the three DL-based in GPU mode can be about 10$\times$ to
32$\times$ faster than in the CPU mode.  It is important to emphasize
that GreenBVQA, as a non-DL-based method, exclusively capitalizes on the
GPU mode for feature generation and regression tasks, resulting in a
noteworthy reduction in inference time, approximately 5$\times$ faster
than its CPU-based counterpart.  On average, GreenBVQA demonstrates an
approximate 1.2x acceleration compared to the second most efficient
method in the GPU mode, namely Mirko \textit{et al.} It is worth noting
that while GreenBVQA may not fully exploit the potential of GPU
acceleration as extensively as DL-based methods, it is foreseeable that
further advancements in third-party libraries and coding optimizations
will yield even more pronounced benefits for GreenBVQA in GPU-supported
environments. 

\begin{figure*}[!htbp]
\centering
\includegraphics[width=0.9\linewidth]{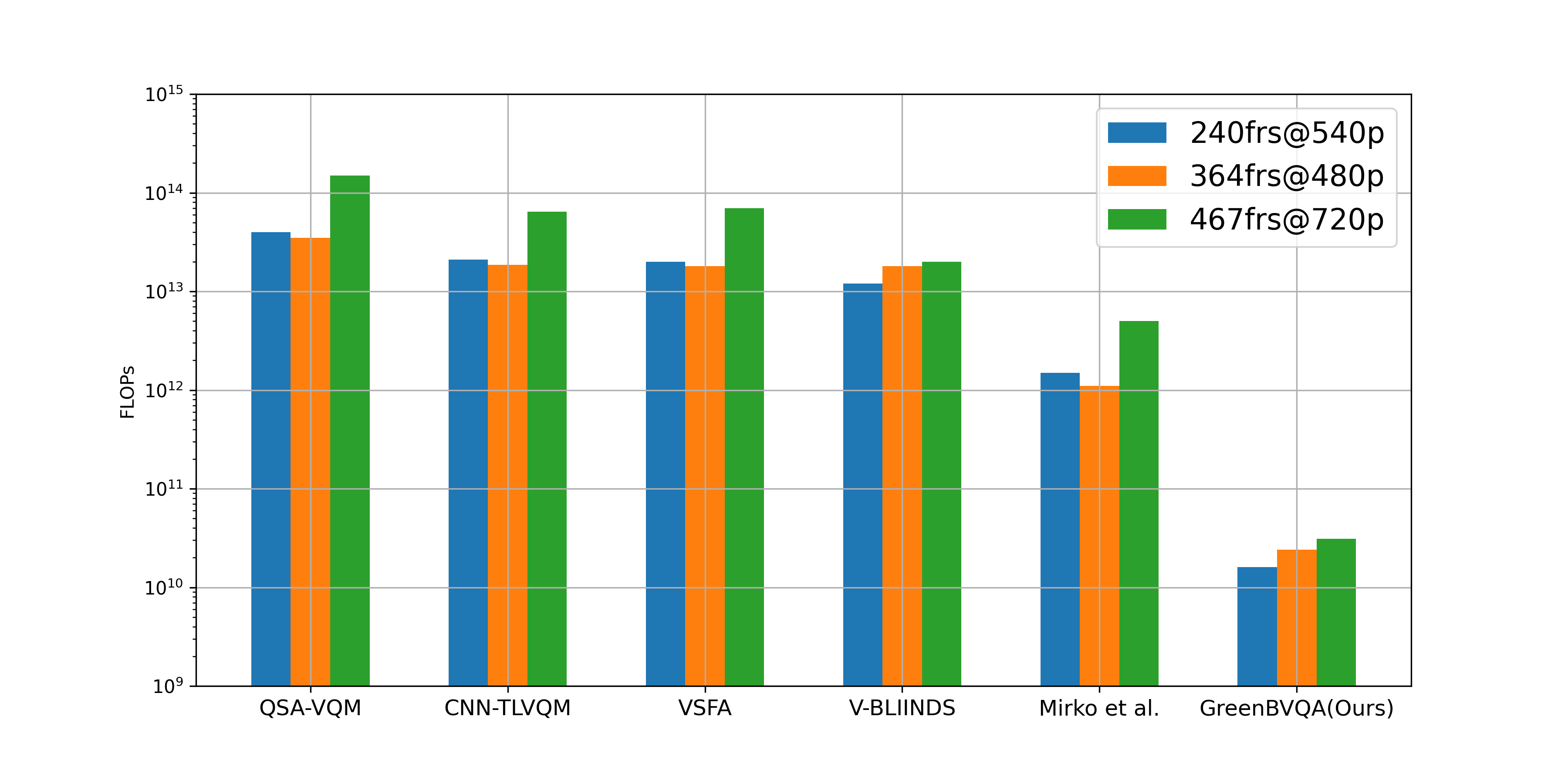}
\caption{Comparison of the floating point operation (FLOP) numbers.}
\label{fig:FLOPs}
\end{figure*}

\subsubsection{Computational complexity}

The number of floating point operations (FLOPs) provides another way to
assess the complexity of a BVQA model.  We estimate the FLOPs of several
DL-based BVQA methods required to predict video quality score and
compare them with that of GreenBVQA.  The FLOPs required by one
240frs@540p test video in the KoNViD-1k dataset are shown in the last
column of Table \ref{table:flop}.  QSA-VQM, CNN-TLVQM, and VSFA demand
remarkably higher FLOPs numbers, ranging from 1250 to 2500 times of
GreenBVQA.  Mirko \textit{et al.}, which is an efficient BVQA method
specifically designed to reduce inference time and computational
complexity, still requires about 100 times of GreenBVQA. 

To be consistent with the inference time analysis, three test videos of
different lengths and resolutions are selected for FLOPs comparison in
Fig.  \ref{fig:FLOPs}.  FLOPS are shown as a function of the frame
number and resolution for several benchmarking methods.  For all the
three videos, QSA-VQM, CNN-TLVQM, VSFA, and V-BLIINDS demand a much
larger number of FLOPs than GreenBVQA. As an efficient method of BVQA,
Mirko \textit{et al.} reduce FLOPs a lot, while still requiring FLOPs of
two orders of magnitude against GreenBVQA.  On the other hand, this
discrepancy is not reflected by the inference time comparison in Table
\ref{table:inference_time} since DL-based methods benifit a lot from
mature coding libraries and environments adept at facilitating deep
learning computations.  The extremetly low FLOPs suggests the potential
for future enhancements in computational efficiency, further bolstering
the appeal of GreenBVQA for real-time video quality assessment tasks in
the edge computing realm. 

\begin{table*}[!htbp]
\centering
\caption{Ablation Study for GreenBVQA.}
\begin{tabular}{l c|c c c c c c c c }
\hline
& & \multicolumn{2}{c}{CVD2014}
& \multicolumn{2}{c}{LIVE-VQC} & \multicolumn{2}{c}{KoNViD-1k}\\\hline
&Model & SRCC & PLCC & SRCC & PLCC & SRCC & PLCC\\\hline
&S features &0.809 &0.844 &0.728 &0.758 &0.720 &0.724\\
&S+T features &0.835 &0.854 &0.762 &0.771 &0.742 &0.749\\ 
&S+T+ST features &- &- &0.776 &0.781 &0.765 &0.766\\
&S+T+ST+SC features &- &- &0.785 &0.789 &0.776 &0.779\\
\hline
\end{tabular}
\label{table:BVQA_ablation}
\end{table*}

\subsection{Abalation Study}

We conduct an ablation study on the choice of selected features in
GreenBVQA. The results are reported in Table \ref{table:BVQA_ablation}.
The examined features include spatial features (S-features), temporal
features (T-features), spatio-temporal features (ST-features), and
spatio-color features (SC-features).  Our study begins with the
assessment of the effectiveness of spatial features (the first row),
followed by adding temporal features (the second row). We see that both
SROCC and PLCC improve in all three datasets.  The addition of
ST-features (the third row) can improve SROCC and PLCC for all datasets
as well.  Finally, we use all four feature types and observe further
improvement in SROCC and PLCC (the last row). Note that the performance
of ST and SC features is not reported for the CVD2014 dataset since
their improvement is little. A combination of S and T features already
reaches high performance for this dataset. 

\begin{figure*}[!htbp]
\centering
\includegraphics[width=1.0\linewidth]{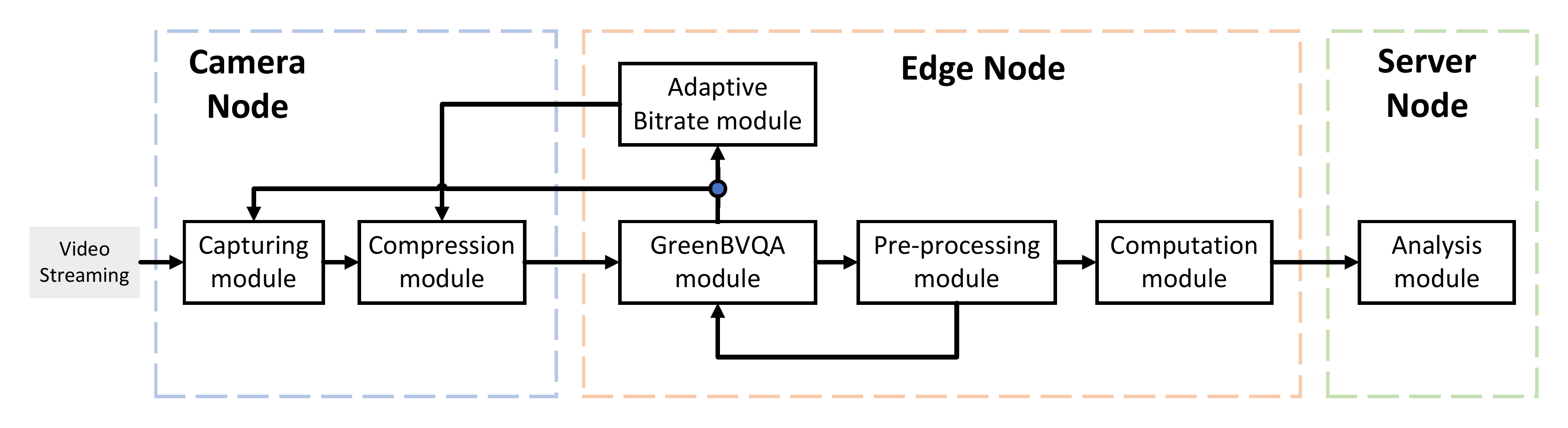}\\
\caption{Illustration of an edge computing system employing GreenBVQA.}
\label{fig:edge_computing}
\end{figure*}

\section{An Edge Computing System with BVQA}\label{sec:edge_computing}

In this section, we introduce a video-based edge computing system to
illustrate the role of GreenBVQA in facilitating various video
processing tasks at the edge. Existing bitrate adaptive video
augmentation methods \cite{sani2017adaptive} primarily consider the
tradeoff between video bitrate and bandwidth consumption to improve the
quality of experience.  Yet, most of them ignore the perceptual quality
of streaming videos.  Perceptual video quality is more relevant to the
human visual experience.  A video with higher bitrates does not
guarantee perceptual friendliness due to the presence of perceptual
distortions such as blurriness, noise, blockiness, etc.  Although the
FR-VQA technique can account for the perceptual quality of streaming
video, the resulting methods rely on reference videos, which are not
available on edge or mobile devices.  As a blind video quality
assessment method, GreenBVQA can operate without any reference.  Its
small model size and low computational complexity make it well-suited
for deployment on edge devices.  Furthermore, its energy efficiency and
cost-effectiveness, evidenced by short inference time, support its
applicability in an edge computing system. 

GreenBVQA can be used as a perceptual quality monitor on edge devices.
An edge computing system that employs GreenBVQA is shown in Fig.
\ref{fig:edge_computing}, where GreenBVQA is used to enhance users'
experience in watching videos.  As shown in the figure, the system
involves predicting the perceptual quality of video streams with no
reference. The predicted quality score can be utilized by other modules
in the system. 
\begin{enumerate}
\item The predicted score can be used as feedback to the phone camera in
video capturing.  In certain extreme situations, such as dark or blurred
video capturing conditions, a low predicted video quality score can
serve as an alert so that the user can change the camera setting to get
improved video quality.
\item In the context of video streaming over the network, it can assist
the adaptive bitrate module in adjusting the bitrate of subsequent video
streams. If the predicted score is higher, the coding module at the
transmitter end can provide a lower bitrate video stream to save the
bandwidth. 
\item Several video pre-processing modules (e.g., video enhancement
\cite{wu2022edge} and video denoising \cite{ge2022edge}) are commonly
implemented on edge devices to alleviate the computational burden of the
server.  By leveraging the predicted video quality scores, unnecessary
pre-processing operations can be saved.  
For instance, when a sequence of video frames is predicted to have a good
visual quality, there is no need to denoise or deblur the frame sequence. 
GreenBVQA can also be used to evaluate the performance of
video pre-processing tasks. 
\end{enumerate}

\section{Conclusion and Future Work}\label{sec:conclusion}

As the demand for high-quality videos captured and consumed at the edge
continues to grow, there is an urgent need for a perceptual video
quality prediction model that can guide these tasks effectively.  
A lightweight blind video quality assessment method called GreenBVQA was
proposed in this work. Its SROCC and PLCC prediction performance was
evaluated on three popular video quality assessment datasets.  
GreenBVQA outperforms all conventional (non-DL-based) BVQA methods and achieves
comparable performance with state-of-the-art (DL-based) BVQA methods.
GreenBVQA's small model size and low computational complexity, which
implies high energy efficiency, make it well-suited for integration into
an edge-based video system. GreenBVQA exhibits short inference times,
enabling real-time prediction of perceptual video quality scores using either CPUs or GPUs. 

There are several promising research directions worth future
investigation. First, in the context of high frame rate video capturing
and transmission, it is desired to adopt adaptive bitrate (ABR) video
with variable frame rates (VFR). With the emergence of advanced edge
devices capable of capturing high frame rate videos, we expect GreenBVQA
to support VFR video and enable ABR video transmission with proper
adaptation. Second, as user-generated content (UGC) videos with diverse
content grow, we see a need to tailor GreenBVQA to specific content
types such as gaming and virtual reality (VR). 

\bibliographystyle{unsrt}  
\bibliography{references}

\end{document}